\documentstyle[aps,graphicx,epsf]{revtex}\tighten
\draft

\def \D {\mbox{D}}

\begin{document}

\title{No Chaos in Brane-World Cosmology}

\author{A. A. Coley$^{1,*}$} \address{$^1$Department of Mathematics and Statistics,
Dalhousie University, Halifax, Nova Scotia}
\maketitle

\begin{abstract}

We discuss the asymptotic dynamical evolution of spatially homogeneous 
brane-world cosmological models close to the initial singularity.
We find that generically  the cosmological singularity is isotropic
in Bianchi type IX brane-world models and consequently these models
do not exhibit Mixmaster or chaotic-like behaviour 
close to the initial singularity.
We argue that this is typical of more general cosmological models in the brane-world scenario.
In particular, we show that an isotropic singularity is a past-attractor in all 
orthogonal Bianchi models
and is a local past-attractor in a class of inhomogeneous brane-world models.

\end{abstract}

 \vskip2pc

\section{Introduction}
Higher-dimensional gravity theories inspired by string
theory, in which the matter fields are confined to a 3-dimensional
`brane-world' embedded in $1+3+d$
dimensions while the gravitational field can also propagate in
the $d$ extra dimensions (i.e., in the `bulk')~\cite{rubakov}, are currently of great interest. 
The $d$ extra dimensions need not be small or even
compact in these theories, which effectively become
4-dimensional at lower energies.
In recent work Randall and Sundrum~\cite{randall} have shown that for
$d=1$, gravity can be localized on a single 3-brane even when the
fifth dimension is infinite. 
In this paradigm,  
Einstein's theory of general relativity (GR) must be modified at high 
energies (i.e., at early times).
An elegant
geometric formulation and generalization of the Randall-Sundrum
scenario has been given~\cite{sms,Maartens}.
Much effort is has already been devoted to
understand the cosmology of such a Randall-Sundrum-type brane-world scenario~\cite{cline,BinDefLan:2000}.

The dynamical
equations on the 3-brane differ from the GR
equations by terms that carry the effects of imbedding and of the
free gravitational field in the five-dimensional bulk~\cite{sms,Maartens}. 
The local (quadratic) energy-momentum corrections are
significant only at very high energies. 
In addition to  the matter
fields corrections, there are nonlocal effects 
from the free gravitational field in the bulk, transmitted via the
projection ${\cal E}_{\mu\nu}$ of the bulk Weyl tensor, that contribute further corrections to the Einstein
equations (e.g., the Friedmann equation) on the brane. Due to its symmetry properties,
${\cal E}_{\mu\nu}$ can be irreducibly decomposed (with
respect to a timelike congruence $u^\mu$) into
\begin{equation}
{\cal
E}_{\mu\nu}=-\left({\widetilde{\kappa}\over\kappa}\right)^4\left[{\cal
U}\left(u_\mu u_\nu+{\textstyle {1\over3}} h_{\mu\nu}\right)+{\cal
P}_{\mu\nu}+2{\cal Q}_{(\mu}u_{\nu)}\right]\,, \label{1}
\end{equation}
in terms of
an effective nonlocal energy density on the brane, ${\cal U}$, arising from the free gravitational field in the
bulk, an effective nonlocal anisotropic stress on the brane,
${\cal P}_{\mu\nu}$, and an effective nonlocal energy flux on the brane,
${\cal Q}_\mu$~\cite{Maartens}.

In general, the 
conservation equations do not determine all of the independent components
of ${\cal E}_{\mu\nu}$ on the brane. In particular, there is no
evolution equation for ${\cal P}_{\mu\nu}$ and hence, in general, the projection of the
5-dimensional field equations onto the brane does not lead to a
closed system. 
However, in the cosmological context,  in which the background metric is
spatially homogeneous and isotropic, we have that
\begin{equation}\label{6c}
\D_\mu{\cal U}={\cal Q}_{\mu}={\cal P}_{\mu\nu}=0\,,
\end{equation}
where $\D_\mu$ is the totally projected part of the brane
covariant derivative.
Since ${\cal P}_{\mu\nu}=0$,
in this case the evolution of ${\cal E}_{\mu\nu}$ is fully determined~\cite{randall}.
In general ${\cal U}={\cal U}(t)\neq0$ (and can be negative) in the Friedmann
background~\cite{cline,BinDefLan:2000}.
For a spatially homogeneous and isotropic
model on the brane equation (\ref{6c}) follows, and 
similar conditions apply self-consistently in other Bianchi models~\cite{Maartens}.
The Friedmann brane-world models have been
extensively investigated \cite{cline,BinDefLan:2000}.

There are many reasons to consider the classical dynamical evolution 
in more general spatially homogeneous Randall-Sundrum-type brane-world cosmological models,
particularly in the early Universe in which the dynamical behaviour can be completely different
to that of GR.
In this article we shall consider the local 
dynamical behaviour of the general Bianchi type IX models close to the singularity.

Due to the existence of monotone functions, it is known
that there are no periodic or recurrent orbits in orthogonal spatially homogeneous
Bianchi type IX models in GR. 
In particular, there are no sources or sinks 
and generically Bianchi
type  IX  models have an oscillatory behaviour with chaotic-like
characteristics, with the matter density becoming dynamically negligible as one
follows the evolution into the past towards the initial singularity.
Using qualitative techniques,  Ma and Wainwright (see \cite{WE}) have 
shown that the orbits of the associated cosmological dynamical system are
negatively asymptotic to a lower two--dimensional attractor. This is the union
of three ellipsoids in ${\bf R}^5$ consisting of the Kasner ring joined by Taub
separatrices; the orbits spend most of the time near the  self-similar Kasner vacuum 
equilibrium points. More rigorous global results are possible. 
Ringstr\"om has proven that a curvature invariant is unbounded in the
incomplete directions of inextendible null geodesics for generic vacuum
Bianchi models, and has rigorously shown that the Mixmaster attractor
is the past attractor of Bianchi type IX models with an orthogonal perfect
fluid \cite{ringstrom}. 

All spatially homogeneous models in GR expand indefinitely except for the Bianchi type IX models.  
Bianchi type IX models
obey the ``closed universe recollapse'' conjecture
 \cite{LW}, whereby initially expanding models enter a
contracting phase and recollapse to a future ``Big Crunch''.  All orbits in the Bianchi IX invariant
sets  are
positively departing; in order to analyse the future asymptotic states of
such models it is necessary to compactify phase-space.  The description of
these models in terms of conventional Hubble- or expansion-normalized variables is only
valid up to the point of maximum expansion (where $H = 0)$.
 An appropriate set of alternative normalised variables, 
which leads to the compactification of Bianchi IX state space, were suggested in \cite{WE}  
(see section 8.5.2)
and have been utilized in the qualitative study of
locally rotationally symmetric perfect fluid models
\cite{Uggla-zurMuhlen}, of scalar field models with an
exponential potential \cite{HO}, and scalar
field models with a perfect fluid \cite{CG}.
In particular, spatially homogeneous Bianchi models containing a scalar
field $\phi$ with an exponential potential of the form
$V(\phi)=\Lambda\exp{k\phi}$ were studied in \cite{IbanezvandenHoogenColey}.
For Bianchi type IX models, if the
parameter $k<\sqrt{2}$ then there exists a set of ever-expanding anisotropic
Bianchi type IX models that isotropize and inflate towards an expanding
power-law inflationary solution, except for
the subset that recollapse \cite{Kitada},  and if $k>\sqrt{2}$ then initially expanding
Bianchi type IX models do not isotropize towards an ever-expanding isotropic
model~\cite{HO}.

\section{Governing equations}

The field equations induced on the brane are derived via an
elegant geometric approach by Shiromizu et al. \cite{sms,Maartens}, using
the Gauss-Codazzi equations, matching conditions and $Z_2$
symmetry. The result is a modification of the standard Einstein
equations, with the new terms carrying bulk effects onto the
brane:
\begin{equation}
G_{\mu\nu}=-\Lambda g_{\mu\nu}+\kappa^2
T_{\mu\nu}+\widetilde{\kappa}^4S_{\mu\nu} - {\cal E}_{\mu\nu}\,,
\label{2}
\end{equation}
where 
\begin{equation}
\kappa^2=8\pi/M_{\rm p}^2\,,\lambda=6{\kappa^2\over\widetilde\kappa^4} \,, ~~ \Lambda =
{4\pi\over \widetilde{M}_{\rm p}^3}\left[\widetilde{\Lambda}+
\left({4\pi\over 3\widetilde{M}_{\rm
p}^{\,3}}\right)\lambda^2\right]\,. \label{3}
\end{equation}

The brane energy-momentum
tensor for a perfect fluid or a minimally-coupled scalar field is given by
\begin{equation}
T_{\mu\nu}=\rho u_\mu
u_\nu+ph_{\mu\nu}\,.
 \label{4}
\end{equation}
where $u^\mu$ is the 4-velocity, $\rho$ and $p$ are the energy density and isotropic
pressure, and the projection tensor $h_{\mu\nu} \equiv g_{\mu\nu}+u_\mu u_\nu$ projects
orthogonal to $u^\mu$.
We shall assume that  the matter content is equivalent
to that of a non-tilting perfect fluid with a linear barotropic
equation of state for the fluid, i.e., 
$p = (\gamma-1)\rho$, where the energy conditions  impose the
restriction $\rho\geq 0$, and the constant $\gamma$ satisfies $\gamma\in[0,2]$
from causality requirements. 
For a minimally coupled scalar field the energy density and pressure are, respectively,
\begin{equation}
\rho={1\over2}\dot\phi^2+ V(\phi),~ p={1\over2}\dot\phi^2-V(\phi),
\end{equation}
and the conservation law is equivalent to the Klein-Gordon equation.
A dynamical analysis of scalar
field models indicates that at early times the scalar field is effectively massless \cite{CG}.
A massless scalar field is equivalent to a perfect fluid
with a stiff equation of state parameter  $\gamma=2$. In the early Universe, and close to
the singularity,  we
expect that $\gamma \geq 4/3$.

The bulk corrections to the Einstein equations on the brane are of
two forms: firstly, the matter fields contribute local quadratic
energy-momentum corrections via the tensor $S_{\mu\nu}$, and
secondly, there are nonlocal effects from the free gravitational
field in the bulk, transmitted via the projection ${\cal
E}_{\mu\nu}$ of the bulk Weyl tensor. The matter corrections are
given by
\[
S_{\mu\nu}={\textstyle{1\over12}}T_\alpha{}^\alpha T_{\mu\nu}
-{\textstyle{1\over4}}T_{\mu\alpha}T^\alpha{}_\nu+
{\textstyle{1\over24}}g_{\mu\nu} \left[3 T_{\alpha\beta}
T^{\alpha\beta}-\left(T_\alpha{}^\alpha\right)^2 \right]\,,
\]
which is equivalent to
\begin{equation}
S_{\mu\nu}={\textstyle{1\over12}}\rho^2
u_\mu u_\nu
+{\textstyle{1\over12}}\rho\left(\rho+2 p\right)h_{\mu\nu}\,,\label{7}
\end{equation}
for a perfect fluid or minimally-coupled scalar field. The quadratic energy-momentum
corrections to standard GR will be significant for
$\widetilde{\kappa}^4\rho^2 \gtrsim \kappa^2\rho$ in the
high-energy regime.

All of the bulk corrections may be consolidated into an effective total
energy density, pressure, anisotropic stress and energy flux, as
follows. The modified Einstein equations take the standard
Einstein form with a redefined energy-momentum tensor:
\begin{equation}
G_{\mu\nu}=\kappa^2 T^{\rm tot}_{\mu\nu}\,,
\label{8}
\end{equation}
where
\begin{equation}
T^{\rm tot}_{\mu\nu} \equiv T_{\mu\nu}+{\widetilde{\kappa}^{4}\over
\kappa^2}S_{\mu\nu}- {1\over\kappa^2}{\cal E}_{\mu\nu}-{\Lambda\over\kappa^2} g_{\mu\nu}\,
\label{9}
\end{equation}
is the redefined perfect fluid (or minimally coupled scalar field) energy-momentum tensor
with
\begin{eqnarray}
\rho^{\rm tot} = \rho+{\widetilde{\kappa}^{4}\over
\kappa^6}\left[{\kappa^4\over12}\rho^2 +{\cal U}\right] + {\Lambda\over\kappa^2}\label{a
}\\
p^{\rm tot} = p+ {\widetilde{\kappa}^{4}\over
\kappa^6}\left[{\kappa^4\over12}(\rho^2+2\rho p)
+{\textstyle{1\over
3}}{\cal U}\right] - {\Lambda\over\kappa^2}\label{b}
\end{eqnarray}
where we recall that
$D_\mu{\cal U}={\cal Q}_{\mu}={\cal P}_{\mu\nu}=0$
in the cosmological case of interest here.

As a consequence of the form of the bulk energy-momentum tensor
 and of $Z_2$ symmetry, it follows \cite{sms} that
the brane energy-momentum tensor separately satisfies the
conservation equations, i.e.,
\begin{equation}\label{12}
\nabla^\nu T_{\mu\nu}=0 \,,
\end{equation}
whence  the Bianchi identities on the brane imply that the projected
Weyl tensor obeys the constraint
\begin{equation}
\nabla^\mu{\cal E}_{\mu\nu}=\widetilde{\kappa}^4\nabla^\mu
S_{\mu\nu}\,. \label{5}
\end{equation}
In the spatially homogeneous case, the conservation equations
for a non-tilting perfect fluid with a linear barotropic
equation of state (or massless scalar field)
reduce to
\begin{equation}
\dot{\rho}+ 3H(\rho+p)=0\,,\label{pc}
\end{equation}
where the Hubble parameter
$H \equiv {\dot a}/a $ gives the volume
expansion rate, and
the nonlocal conservation equations 
reduce to an evolution equation for ${\cal U}$:
\begin{equation}
\dot{\cal U}+ 4H{\cal U}=0\,.
\label{pc'}
\end{equation}
In the Friedmann
case  Eq.
(\ref{pc'}) yields the `dark radiation' solution
${\cal U}={\cal U}_o\left({a_o/a}\right)^4$.

The field equations are given by Eqns (\ref{8}) and (\ref{9}).
The generalized Friedmann equation, which determines the expansion of the universe
or the Hubble function, in the case of spatially homogeneous
cosmological models is
\begin{equation} 
H^2 = \frac{1}{3}\kappa^2\rho\left(1+\frac{\rho}{2\lambda}\right)
-\frac{1}{6}{}^3R+\frac{1}{3}\sigma^2+\frac{1}{3}\Lambda +
\frac{2{\cal U}}{\lambda\kappa^2}\,, \label{frie}
\end{equation}
where ${}^3R$ is the scalar curvature of the hypersurfaces orthogonal
to the fluid flow, which we associate with the cosmological
fluid, and  $2\sigma^2\equiv\sigma^{ab}\sigma_{ab}$ is the
shear scalar in terms of the shear $\sigma_{\mu\nu}$.

For a perfect fluid or minimally-coupled scalar field, the
generalized Raychaudhuri equation on the brane becomes~\cite{Maartens}:
\begin{equation}
\dot{H}+ H^2+  {2\over3} \sigma^2
+{\textstyle{1\over6}}\kappa^2(\rho + 3p) -{1\over3}\Lambda 
= -{1\over36}\left({\widetilde{\kappa}\over\kappa}\right)^4 \left[
{\kappa}^4 \rho(2\rho+3p)
 +12{\cal U}\right], \label{pr}
\end{equation}
where we have assumed (\ref{6c}), and in particular
${\cal P}_{\mu\nu}=0$,
self-consistently on the brane~\cite{Maartens}.

\section{Bianchi IX Models}

Following \cite{WE,CamSop} we define Hubble-normalized (and dimensionless) shear variables
$\Sigma_+,\Sigma_-$, curvature variables $N_1,N_2,N_3$  and matter variables $\Omega_i, P_i$
(relative to a group-invariant orthonormal frame),
and a logarithmic (dimensionless) time variable, $\tau$, defined by $d\tau = H dt$.
These variables do not lead to a global compact phase space, but they are
bounded close to the singularity \cite{WE}. 
The governing evolution  equations for these quantities are then
\begin{eqnarray}
 \Sigma_+' = (q-2)\Sigma_+ -S_+\\
 \Sigma_-' = ( q-2)\Sigma_- - S_-\\
  N_1 '     = ( q -4\Sigma_+) N_1\\
  N_2 '     =  (q+2\Sigma_++2\sqrt{3}\Sigma_-)N_2\\
  N_3 '     =  (q +2\Sigma_+-2\sqrt{3}\Sigma_-)N_3,
\end{eqnarray}
where a prime denotes differentiation with respect to $\tau$, and $K$, $S_+$ and
$S_-$ are curvature terms that are defined as follows:

\begin{mathletters}
\begin{eqnarray}
{K}&\equiv&  \frac{1}{12}\left((N_1^2+N_2^2+N_3^2)
        -2( N_1N_2+N_2N_3+ N_1N_3)\right)\\
{{S}_{+}}&\equiv&{\displaystyle \frac {1}{6}}\, \left( \! \,
        {{N}_{2}} - {{N}_{3}}\, \!  \right) ^{2}
        - {\displaystyle \frac {1}{6}}\,{{N}_{1}}\,
        \left( \! \,2\,{ {N}_{1}} - {{N}_{2}} -
        {{N}_{3}}\, \!  \right) \\
{{S}_{-}}&\equiv& \frac{1}{6}\,\sqrt {3}\, \left( \!
\,{{N}_{2}} -
         {{N}_{3}}\, \!  \right) \, \left( \! \,
         - {{N}_{1}} + {{N}_{2}} + {{N}_{3}}\, \!  \right)
\end{eqnarray}
\end{mathletters}
The quantity $q$ is the deceleration parameter given by
\begin{equation}
q \equiv   2\Sigma_+^2+2\Sigma_-^2  + \frac{1}{2}\sum \Omega_i + \frac{3}{2}\sum P_i.\label{q}
\end{equation}
The decoupled Raychaudhuri equation becomes
\begin{equation}
H '     = -(1 + q)H.\label{raych}
\end{equation}
In addition, the generalized Friedmann equation reduces to the constraint 
\begin{equation}     
1 =  \Sigma_+^2+
\Sigma_-^2+ \sum \Omega_i + K. \label{FRIEM}
\end{equation}

Due to the symmetries in the dynamical system, we can restrict ourselves to
the set  $ N_1\geq 0$, $N_2\geq 0$, and
$N_3\geq 0$, without loss of generality.  (Note, for Bianchi IX
models all of the $N_i$'s must be of the same
sign.) We again note that these new normalized variables, for which the evolution equation for
$H$ has decoupled from the remaining equations, are bounded close to the initial
singularity.

In the above
\begin{equation}
 \sum \Omega_i =  \frac{\kappa^2 \rho^{tot}}{3H^2},~~
 \sum P_i =  \frac{\kappa^2 P^{tot}}{3H^2},  \label{def}
\end{equation}
where the $\Omega_i$ ($i= 1-4$) are given by
\begin{mathletters}
\begin{eqnarray}
 \Omega_1 =  \Omega = \frac{\kappa^2 \rho}{3H^2}; ~ P = (\gamma -1)\Omega \label{d1}\\
 \Omega_2 =  \Omega_b =  \frac{\kappa^2 \rho^2}{6 \lambda H^2}; ~ P_b = (2\gamma -1)\Omega_b\label{d2}\\
 \Omega_3 =  \Omega_{\cal U} =  \frac{2 {\cal U}}{\kappa^2 \lambda H^2}; ~ P_{\cal U} = \frac{1}{3}\Omega_{\cal U}\label{d3}\\
 \Omega_4 = \Omega_\Lambda =  \frac{\Lambda }{3 H^2}; ~ P_\Lambda = -\Omega_\Lambda, \label{d4}
\end{eqnarray}
\end{mathletters}
which satisfy the equations
\begin{equation}
\Omega_i ' = [2(1+q) -3 \Gamma_i] \Omega_i,\label{CE}
\end{equation}
where
\begin{equation}
\Gamma_1=\gamma, \Gamma_2 = \Gamma_b = 2 \gamma, \Gamma_3 = \Gamma_{\cal U} = \frac{4}{3}, 
\Gamma_4= \Gamma_\Lambda = 0.
\end{equation}

A minimally coupled homogeneous scalar field $\phi=\phi(t)$ with an exponential
potential can also be included,  where
the energy-momentum tensor is given by
\begin{equation}
T_{ab}^{\ \ sf} =
\phi_{;a}\phi_{;b}-g_{ab}\left(\frac{1}{2}\phi_{;c}\phi^{;c}+V(\phi)\right).
\end{equation}
The usual (linear) terms can be included via \cite{CG}
\begin{eqnarray}
 \Omega_\phi = \Psi^2 + \Phi,
 P_ \phi =  \Psi^2 - \Phi, \label{defsf}
\end{eqnarray}
where
\begin{equation}
\Phi\equiv\frac{\kappa^2 V}{3H^2},\qquad\qquad
\Psi\equiv\frac{\kappa^2\dot\phi}{\sqrt{6}H}\label{dv1}
\end{equation}
The evolution of these variables is then given by the conservation or  Klein-Gordon equation,
which can be written as
\begin{eqnarray}
  \Psi '     = (q -2)\Psi
                                       -\frac{\sqrt{6}}{2}k\Phi^2 \\
 \Phi '    =\Phi\left((1+ q)
                +\frac{\sqrt{6}}{2}k\Psi\right)
\end{eqnarray}
The local quadratic corrections 
$\Omega_{b\phi}$ and $P_{b\phi}$ can be included via
$\Phi_b \propto \frac{V}{H}$ and
$\Psi_b \propto \frac{\dot\phi}{H^2}$ and a generalized (quadratic)
Klein-Gordon equation.
Since an extensive analysis of scalar field models has shown that close to the
initial singularity the scalar field must be massless \cite{CG,IbanezvandenHoogenColey}
(see also \cite{R2000,Foster}), it is plausible that
scalar field models can be approximated by a stiff perfect fluid 
close to the
initial singularity (particularly regarding questions of stability).

\subsection{Initial Singularity}

If $\rho^{tot} > 0$ and $\rho^{tot} + 3 p^{tot}> 0$ (which follows from the various energy conditions)
for all $t < t_0$, then from the generalized  Raychaudhuri equation (\ref{pr})
(or (\ref{raych})) and using the generalized Friedmann equation (\ref{frie})
(or (\ref{FRIEM})) and the conservation equations, it follows that
for $\dot {a_0} > 0$ (where $a_0 \equiv a(t_0)$) there exists a time $t_b$ with
$t_b<t_0$ such that $a(t_b)=0$, and there exists a singularity at $t_b$, where
we can rescale time so that $t_b=0$ and the singularity occurs at the origin.

From Eqn (\ref{pr}) we can find the precise constraints on $\Lambda$ and ${\cal U}_0$
in terms of $a_0$ in order for these conditions to be satisfied at $t = t_0$. It then follows from
the generalized Raychaudhuri equation, the generalized Friedmann equation 
and the conservation equations that if these conditions are satisfied at $t = t_0$
they are satisfied for all $0 < t < t_0$, and a singularity necessarily results.
These conditions are indeed satisfied for regular matter undergoing thermal collapse in which the 
the local energy density and pressure satisfy
$\rho(2\rho+3p)>0$ (and is certainly satisfied for perfect fluid matter satisfying 
the weak energy condition $\rho \geq 0$ and a
linear barotropic equation of state with $\gamma \geq 1$). 
On the other hand, it is known that a large positive cosmological constant $\Lambda$ or
a significant negative  nonlocal term ${\cal U}$ 
counteracts gravitational collapse and can lead to the singularity being avoided 
in exceptional circumstances.

Since the variables are bounded  close to the singularity,
it follows from Eqn (\ref{q}) that $0< q < 2$.
Hence, from Eqn (\ref{raych}), $H$ diverges as the initial singularity is approached.
At an equilibrium point $q = q^*$, where $q^*$ is a constant with $0< q^* < 2$,
so that from Eqn (\ref{raych}) we have that $H  \rightarrow (1 + q^*)^{-1} t^{-1} $
as $ t  \rightarrow 0^{+}$ ($\tau  \rightarrow -\infty$).

From Eqns (\ref{q}), (\ref{FRIEM}) and the conservation laws
it then follows that $\rho \rightarrow \infty$
as $ t  \rightarrow 0^{+}$. It then follows directly from the
conservation laws (\ref{CE}) that $\Omega_b$ dominates as $ t  \rightarrow 0^{+}$
and that all of the other $\Omega_i$ are negligable dynamically as the singularity
is approached.  The fact that the effective equation of state
at high densities become ultra stiff, so that the 
matter can dominate the shear dynamically, is a unique feature of brane cosmology.

Note that the function defined in Bianchi IX models by
\begin{equation}
Z \equiv (N_1 N_2 N_3)^2, 
\end{equation}
satisfies the evolution equation
\begin{equation}
Z ' = 6qZ, 
\end{equation}
and is consequently monotone close to the singularity~\cite{WE}.

In summary, generically the Bianchi type IX models have a cosmological initial singularity
in which $\rho \rightarrow \infty$, and consequently $\Omega_b$ dominates, as $ t  \rightarrow 0^{+}$.
This can be proven by more rigorous methods~\cite{ren2,ringstrom}. It can also be shown by qualitative methods
that the  spatial 3-curvature is negligable at the initial singularity and from a
comprehensive analysis that at
later times $\rho$ decreases and the low density regime in which GR is valid ensues.

\subsection{The Isotropic Equilibrium Point}

Hence we have shown that close to the singularity 
$\rho^{tot} = \rho_b$, so that  $\sum \Omega_i = \Omega_b$ and
$\sum P_i = P_b$, where $P_b = (2\gamma -1)\Omega_b$. Consequently, Eqn (\ref{FRIEM})
can be written as
\begin{equation}     
\Omega_b = 1-\Sigma_+^2 - \Sigma_-^2 - K,
\end{equation}
which can be used to eliminate $\Omega_b$ from the governing equations. In particular,
Eqn (\ref{q}) becomes
\begin{equation}
q = 3(1-\gamma)(\Sigma_+^2+\Sigma_-^2) + (3\gamma - 1)(1 - K) , 
\end{equation}
and the governing equations are given by the dynamical system (18-22), where
 $q$ now given by the above expression.

There is an equilibrium point of the dynamical system, 
denoted by ${\cal F}_b$, given by $\Sigma_+ = \Sigma_- = 0$, and $ N_1= N_2 = N_3 = 0$,
which corresponds to spatially homogeneous and isotropic non-general-relativistic brane-world
(without brane tension; ${\cal U} = 0$)
models, first discussed by Bin\'etruy, Deffayet and 
Langlois~\cite{BinDefLan:2000}, in which $a(t)\sim t^{\frac{1}{3\gamma}}$,
(which is valid at very high 
energies ($\rho \gg \lambda$) as the initial singularity is approached; $t \rightarrow 0$).
Note that these solutions are self-similar, and are  referred to as
Friedmann brane-worlds \cite{cline}, Bin\'etruy, Deffayet and 
Langlois solutions \cite{BinDefLan:2000}\, or
Brane-Robertson-Walker models~\cite{COLEY}.

For the equilibrium point ${\cal F}_b$, the 5 eigenvalues are:
\begin{equation}
3(\gamma - 1), 3(\gamma - 1), (3\gamma -1), (3\gamma -1), (3\gamma -1)
\end{equation}
Hence, for all physically relevant values of $\gamma$ ($\gamma \geq 1$),
${\cal F}_b$ is a source (or past-attractor) in the brane-world scenario and 
the singularity is isotropic.
This contrasts to the situation in GR in which 
anisotropy dominates for $\gamma<2$.
This is also consistent with previous analyses of Bianchi type I and V models where
${\cal F}_b$ is always a source for $\gamma \ge 1$ 
(in the FRW models, ${\cal F}_b$ is a
source  when $\gamma \geq 1/3$ when ${\cal U} = 0$ and $\gamma \geq 2/3$ when
${\cal U} \ne 0$)~\cite{CamSop}.

\subsection{Discussion}

There are no other equilibrium points of the dynamical system that correspond to sources. However, it is instructive
to consider the Kasner equilibrium points. 

The one-parameter set (circle) of (Bianchi I, shearing) Kasner vacuum ($\Omega_i = 0$)
 equilibrium points ${\cal K}$ are given by $\Sigma_+^2 +  \Sigma_- ^2 = 1$, and $ N_1= N_2 = N_3 = 0$.
 The eigenvalues are 
\begin{equation}
2(1 - 2\Sigma_+), 2(1 + \Sigma_+ + \sqrt{3}\Sigma_-), 2(1 + \Sigma_+ - \sqrt{3}\Sigma_-), -6(\gamma -1), 
\end{equation}
and a fifth eigenvalue which is zero due to the fact that ${\cal K}$ is a one-parameter set 
of equilibrium points. All of these equilibrium points are saddles; in particular, they can never be sources.
The fourth eigenvalue is negative, whereas in GR it is positive, so
that the structure of the stable and unstable manifolds close to ${\cal K}$ is altered.
In GR it was argued \cite{WE} that $\Omega$ and $\Delta$, defined by
\begin{equation}
\Delta \equiv (N_1 N_2)^2 +(N_2 N_3)^2 + (N_3 N_1)^2,
\end{equation}
satisfy $\Omega \rightarrow 0$, $\Delta \rightarrow 0$ as $\tau \rightarrow -\infty$,
and hence in general all orbits approach  ${\cal K}$ 
(which is characterized by $\Omega = 0$, $\Delta = 0$)
into the past. A qualitative and
numerical investigation indicates that $\Omega$ does not tend to zero as 
 $\tau \rightarrow -\infty$, and hence generically  ${\cal K}$ 
is not, nor does it constitute part of, a (past-) attracting set in the models under investigation.

Close to the singularity we expect that $\gamma \geq 4/3$. However, we note that there is a bifurcation 
at $\gamma = 1$, and ${\cal F}_b$ and ${\cal K}$ coalesce to form a 
two-parameter set of equilibrium points ${\cal J}$ which are 
analogues of the Jacobs stiff fluid solution in GR in which $\Omega_b \ne 0$. A subset of these 
equilibrium points are sources. In general $\Sigma_+^2 +  \Sigma_- ^2 \ne 0$ (the equilibrium points
with zero shear are the special points in ${\cal J}$ equivalent to ${\cal F}_b$); however, there are sources 
in ${\cal J}$ in which $\Sigma_+^2 +  \Sigma_- ^2$ (and hence the shear) is arbitrarily small.

\section{Conclusions}

Therefore,  generically the Bianchi IX brane-world models do not have space--like and
oscillatory  singularities in the past, and consequently brane-world cosmological models
do not exhibit Mixmaster and chaotic-like behaviour 
close to the initial singularity.

We expect this to be a generic feature of more general cosmological models in the brane-world scenario.
In particular, we anticipate that ${\cal F}_b$ is a  source in all spatially homogeneous models.
Following arguments similar to those of section III.A, it follows that there exists a
singularity of a similar nature in all orthogonal Bianchi brane-world models. In general,
$\Omega_b$ will again dominate as the initial singularity is approached into the past
and the qualitative results of section III.B will follow. In particular, ${\cal F}_b$
will be a local source and in general the initial singularity will be isotropic.


\subsection{Inhomogeneous Brane-World Models}

Let us consider the dynamics
of a class of spatially inhomogeneous cosmological models with one spatial
degree of freedom in the brane-world scenario. The $G_{2}$ cosmological models
admit a 2-parameter Abelian isometry group
acting transitively on spacelike 2-surfaces. These models  admit one
degree of freedom as regards spatial inhomogeneity, and the resulting governing
system of evolution equations constitute a system of autonomous partial differential
equations in two independent variables.
We follow the formalism of
\cite{elst} which utilizes 
area expansion normalized scale-invariant
dependent variables, and we use the 
timelike area gauge
to discuss the asymptotic evolution of the class of
orthogonally transitive $G_{2}$ cosmologies near the cosmological
initial singularity.  The 
initial singularity can be shown to be characterized by
$E_{1}{}^{1} \rightarrow  0$ as 
$\tau \rightarrow -\infty$, where $E_{1}{}^{1}$ is a normalized frame variable \cite{elst}.

The `equilibrium point' ${\cal F}_b$ is characterized by zero shear
($\Sigma_- = \Sigma_{\times} = 0$) and zero curvature 
($N_- = N_{\times} = 0$) (where we note that the area expansion normalized variables
utilized in \cite{elst} are equivalent to H-normalized variables
close to ${\cal F}_b$). Linearizing the evolution equations about ${\cal F}_b$,
using the same spatial reparametrisation as in \cite{elst}
(so that $E_{1}{}^{1} = \exp((3\gamma-1)\tau)$), we obtain the following general solution of the 
linearized equations
in which  the shear variables satisfy
\begin{equation}
\Sigma_-  =  a_{1}(x)\exp{(3(\gamma-1)\tau)}, ~~ \Sigma_{\times} =  a_{2}(x)\exp{(3(\gamma-1)\tau)},
\end{equation}
the curvature variables satisfy
\begin{equation}
N_-  =  a_{3}(x)\exp{((3\gamma-1)\tau)},~~ N_{\times}  =  a_{4}(x)\exp{((3\gamma-1)\tau)},
\end{equation}
and 
\begin{equation}
\Omega_b =  1 + a_{5}(x)\exp{(3(\gamma-1)\tau)},
\end{equation}
where the $a_{i}(x)$ are arbitrary functions of the space coordinate. 
As in \cite{elst}, it can be shown that the {\em tilt} or peculiar velocity $v$
between the timelike area gauge and the fluid 4-velocity obeys
$v = a_{6}(x)\exp{((3\gamma-1)\tau)}$ and hence tends to zero as $\tau \rightarrow -\infty$
and, in addition, it follows from the conservation laws that 
$\Omega \rightarrow 0$, $\Omega_{\cal U} \rightarrow 0$  and $\Omega_\Lambda \rightarrow 0$
as the initial singularity is approached.

Strictly speaking, an inhomogeneous energy density is not consistent with 
${\cal Q}_{\mu}=0$, ${\cal P}_{\mu\nu}=0$; in particular, a non-zero
$\D_\mu{\rho}$ acts as a source for ${\cal Q}_{\mu}$.
(Physically  ${\cal P}_{\mu\nu}$ corresponds to graviational waves and
will not affect the the dynamics close to the singularity.)
Writing $\mid {\cal Q}_{\mu} \mid = {\cal Q}$, an analysis of the
evolution equation for ${\cal Q}$ \cite {Maartens} close to ${\cal F}_b$
then yields ${\cal Q}H^{-2} \sim  Q(x)\exp{(2(3\gamma-2)\tau)}$, so that 
${\cal Q}H^{-2} \rightarrow 0$ as $\tau \rightarrow -\infty$ (as the initial
singularity is approached). Moreover, it can then be shown that
a small ${\cal Q}$
does not affect the dynamical evolution of ${\cal U}$ (to lowest order) close 
to  ${\cal F}_b$ (indeed, ${\cal Q}/{\cal U}\sim O(1)$, so that if ${\cal Q}$
is negligible it remains negligible). Hence,  
equations (43) -- (45) are consistent with the evolution of ${\cal Q}_{\mu}$,
and consequently represent a 
self-consistent solution  close to ${\cal F}_b$.

The above linearized solution represents a general
solution in the neighbourhood of the initial singularity. Hence ${\cal F}_b$ is a local source or 
past-attractor 
in  this family of spatially inhomogeneous cosmological models for $\gamma >1$.
In particular, we see that the shear and
curvature asymptote to zero as $\tau \rightarrow -\infty$, and hence the
singularity is isotropic.
We also note that, unlike the analysis of the perfect fluid GR models in \cite{elst},
the Kasner equilibrium set ${\cal K}$ are found to be saddles in the class of $G_{2}$ 
brane-world cosmological models.

The most detailed proposal for the structure of space--time singularities in GR
are the conjectures of Belinskii, Khalatnikov and Lifshitz (BKL) \cite{bkl}, one aspect
of which is that each spatial point evolves towards the singularity as 
if it were a spatially homogeneous (Bianchi) cosmology. 
That, is, generic space--times  have the
property that spatial points decouple near the singularity, 
the EFE effectively reduce to ordinary
differential equations 
(i.e., the spatial derivatives have a negligible effect on the
dynamics) and the local
dynamical behaviour is 
asymptotically like that of Bianchi models near the
singularity.
Support for this conjecture comes from recent analyses that show that
the presence of the
inhomogeneity ceases to govern the dynamics asymptotically toward the
singularity in particular classes of inhomogeneous
models. 
In a special class of
Abelian $G_{2}$ spatially inhomogeneous
models, the 
so-called ``velocity-dominated''
spacetimes, the evolution at different spatial points will approach that of
different Kasner solutions ~\cite{inv}. 
A numerical investigation of a class of vacuum Gowdy $G_{2}$ 
cosmological spacetimes that represent an inhomogeneous generalization of
Bianchi type VI$_0$ models with a magnetic field  
has shown evidence that at a generic point in space the
evolution toward the initial singularity is asymptotically that of
a spatially homogeneous spacetime with Mixmaster behaviour~\cite{WIB}.
More rigorous results are given in \cite{elst,G2}. 

The dynamics of
the class of inhomogeneous brane-world models considered above, together with the
BKL conjectures, indicate that a wide class of inhomogeneous brane-world models will
have an isotropic initial singularity.

\subsection{Discussion}

Therefore, it is plausible that generically the brane-world cosmological models have
an {\em isotropic initial singularity\/},
whose generic evolution near the cosmological initial singularity is 
approximated by a spatially homogeneous and
isotropic model in a rigorously defined mathematical sense~\cite{GW85}
(in contrast to the situation in GR).
Therefore, it is possible that brane cosmology provides for a quiescent~\cite{Barrow} 
initial cosmological period in which the Universe is smooth and highly symmetric,
perhaps explained by entropy arguments and the second law of 
thermodynamics~\cite{Penrose79},
thereby naturally providing the precise
conditions for inflation to  subsequently take place
and consequently avoiding the problem of initial conditions in inflation
\cite{COLEY}.

The results in this paper are incomplete in that
a description of the gravitational field in the
bulk is not provided. Unfortunately, the evolution of the
anisotropic stress part is {\em not} determined on the brane.
These nonlocal terms also enter into crucial dynamical equations,
such as the Raychaudhuri equation and the shear propagation
equation, and can lead to important changes from the 
GR case. 
The correction terms must be consistently derived from the higher-dimensional equations.
Additional modifications can also occur for higher-dimensions (than 5), 
for more general (than static) higher-dimensional (bulk) geometry, 
for  higher (non-GR) curvature corrections,
for higher-dimensional matter fields (e.g., scalar fields) in the bulk
and for motion of the brane~\cite{frolov}. Further work is therefore
necessary. However, it is plausible that the main
results of this paper will persist when these additional effects are included.
In particular, this is supported by a
dynamical analysis of the evolution of ${\cal Q}_{\mu}$ and the fact that since ${\cal P}_{\mu\nu}$
corresponds to gravitational waves in higher-dimensions it is expected that the dynamics
will not be affected significantly at early times close to the singularity.

It is also of importance to examine what happens when anisotropic stresses
are included. Recently, Barrow and Hervik \cite{hervik} studied a class of Bianchi type I brane-world
models with a pure magnetic field and a perfect fluid with a linear barotropic $\gamma$-law equation of state.
They found that when $\gamma \ge \frac{4}{3}$,  the equilibrium point ${\cal F}_b$
is again a local source (past-attractor), but that there exists a second equilibrium point
denoted $PH_1$, which corresponds to a new brane-world solution with a non-trivial
magnetic field, which is also a local source. When $\gamma < \frac{4}{3}$, $PH_1$
is the only local source (the equilibrium point ${\cal F}_b$ is unstable to magnetic field
perturbations and hence is a saddle); however, even in this case it was argued
that chaotic behaviour is not possible \cite{hervik}.
Finally, we note that Bianchi type I and IX models have also been studied in 
Ho\v{r}ava-Witten cosmology in  which the fifth coordinate is a $S^1/Z_2$ orbifold while the remaining
six dimensions have already been compactified on a Calabi-Yau space, and 
it was argued \cite{Dabr} 
that there is no chaotic behaviour in such Bianchi IX  Ho\v{r}ava-Witten cosmologies.

\[ \]
{\bf Acknowledgements:}\\

This work was supported, in part, by NSERC of Canada.

\end{document}